\documentstyle[epsfig,manuscript,aps]{revtex}
\tightenlines
\begin{document}

\title{Tunable spin transport in CrAs: role of correlation effects}
\author{L. Chioncel$^{1,2}$, M.I. Katsnelson$^{1}$, G.A. de Wijs$^{1}$, \\
R.A. de Groot $^{1,3}$, and A. I. Lichtenstein$^{4}$}
\address{$^1$Radboud University Nijmegen, NL-6525 ED Nijmegen, The Netherlands \\
$^2$ Institute for Theoretical Physics and Computational
Physics, Graz University of Technology, A-8010 Graz, Austria \\
$^3$Materials Science Centre, NL-9747 AG Groningen, The Netherlands \\
$^4$ Institute of Theoretical Physics, University of Hamburg, 20355 Hamburg, Germany}
\maketitle

\begin{abstract}
Correlation effects on the electronic structure of half-metallic CrAs in
zinc-blende structure are studied for different substrate lattice constants.
Depending on the substrate the spectral weight of the non-quasiparticle states
might be tuned from a well developed value in the case of InAs substrate to an almost
negligible contribution for the GaAs one. A piezoelectric material that would allow
the change in the substrate lattice parameters opens the possibility for practical
investigations of the switchable (tunable) non-quasiparticle states. Since the
latter are important for the tunneling magnetoresistance and related phenomena
it creates new opportunities in spintronics.
\end{abstract}

\section{Introduction}
One of the strongest motivations to investigate magnetic
semiconductors and half-metallic ferromagnets (HMF) is the
possibility to design and produce novel stable structures on
semiconducting substrates with new interesting properties.
Adopting this point of view first-principle studies are an
excellent starting point to predict new systems having the desired
properties. Recently Akinaga {\it et al.} \cite{Akinaga00} found
the possibility to fabricate zinc-blende (ZB) type CrAs
half-metallic ferromagnetic material. Experimental data confirmed
that this material is ferromagnetic with the magnetic moment of $3
\mu_B$, in agreement with theoretical predictions
\cite{Akinaga00}. According to this calculation this half-metallic
material has a gap of about $1.8eV$ in the minority spin channel
which has attracted much attention to this potential candidate for
spintronic applications, keeping in mind also its high Curie
temperature $T_c$ around $400K$. Note that recent experiments on
CrAs epilayers grown on GaAs$(001)$ evidenced an orthorhombic
structure, different from the ZB one, so the structure is rather
sensitive to the preparation process \cite{Etgens04}.
However, it is highly desirable to explore the possibility of existence
of half-metallic ferromagnetism in materials which are compatible
with important III-V and II-IV semiconductors. For this purpose efforts
has been made on the metastable ZB structures, such as CrAs \cite{Akinaga00,Mizguchi02}.
Beyond the importance of the preparation techniques, attention should be 
given to the understanding of the finite temperature properties in these
HMF materials. Therefore, it is interesting to explore theoretically
the mechanism behind half-metallic ferromagnetism at finite temperature
from a realistic electronic structure point of view.

Theoretical studies \cite{Shirai01} of the $3d$ transition metal
monoarsenides have shown that the ferromagnetic phase of ZB
structure CrAs compound should be more stable than the
antiferromagnetic one. The calculated equilibrium lattice constant
is larger than that of GaAs ($5.65 \AA$) having a value of $5.80
\AA$ \cite{Shirai01}. Following this work similar electronic
structure calculations concerning the stability of the
half-metallic ferromagnetic state in the ZB structure have been
carried out \cite{Xie03}.

However standard local density approximation LDA (or GGA, LDA+U,
etc.) calculations are in general insufficient to describe some
important many-particle features of the half-metallic
ferromagnets. One of these many-body features, the
non-quasiparticle (NQP) states \cite{edwards,IK,ufn} contribute
essentially to the tunneling transport in heterostructures
containing HMF \cite{ourtransport,falko}. The origin of these
states is connected  with the ``spin-polaron'' processes: the
spin-down low-energy electron excitations, which are forbidden for
HMF in the one-particle picture, turn out to be possible as
superpositions of spin-up electron excitations and virtual magnons
\cite{edwards,IK}.  Recently we have applied the LDA+DMFT
(dynamical mean field theory)  approach
\cite{anisDMFT,Katsnelson99} to describe from first principles the
non-quasiparticle states in a prototype half-metallic ferromagnet
NiMnSb \cite{NQP}.

In this paper, we describe the correlation effects in CrAs HMF
material in the framework of the LDA+DMFT approach.  We will show
that these many-body spin-polaron processes are very sensitive to
structural properties  of the artificially produced CrAs compound.
Depending on the substrate characteristics, such as a large
lattice constant, as in the case of InAs ($6.06 \AA$), or a
smaller one as in the case of GaAs ($5.65 \AA$) the  spectral
weight of the non-quasiparticle states can be tuned from a large
value in the former case (InAs) to an almost negligible
contribution in the later case (GaAs). Therefore, the correlation
effects in conjunction with structural properties determine the
behavior of electronic states in CrAs near the Fermi level which
has a substantial impact on the tunneling transport in the
corresponding heterostructures.

\section{Tunneling transport and non-quasiparticle physics}
First, let us explain in  a simple way why the non-quasiparticle
states are important for the tunneling transport (for formal
derivations, see Refs. \onlinecite{ourtransport,falko}). To this
aim we consider the case of a narrow-band saturated Hubbard
ferromagnet where the current carriers are the holes in the lowest
Hubbard band and the non-quasiparticle states provide {\it all}
spectral weight for the minority spin projection \cite{IK}. A
schematic density of states is shown in Fig. \ref{model}a. Suppose
we have a tunnel junction with two pieces of this ferromagnet with
either parallel (Fig. \ref{model}b) or antiparallel (Fig.
\ref{model}c)  magnetization directions. From the one-particle
point of view, the spin-conserving tunneling is forbidden in the
latter case. However,  in the framework of many-particle picture
the charge current is a transfer process between an empty site and
a single-occupied site rather than the motion of the electron
irrespective to the site like in the band theory and therefore the
distinction between these two cases (see Fig. \ref{model}), is due
only to the difference in the densities of states. It means
that the estimations of the tunneling magnetoresistance based on
simple one-electron picture is too optimistic; even for
antiparallel spin orientation of two pieces of the half-metallic
ferromagnets in the junction for zero temperature the current is
not zero, due to the non-quasiparticle states. More exactly, it
vanishes for zero bias since the density of NQP states at the
Fermi energy equals to zero. However, it grows with the bias
sharply, having the scale of order of typical {\it magnon} energies,
that is, millivolts.

The latter statement is confirmed by a formal consideration of the 
antiparallel spin orientation case, based 
on the standard tunneling Hamiltonian \cite{Mahan90}:
\begin{equation}
H=H_L+H_R+\sum_{kp}(T_{kp}c_{k\uparrow}^{\dagger}c_{p\downarrow} + h.c.)
\end{equation}
where $H_{L,R}$ are the Hamiltonians of the left (right)
half-spaces, $k$ and $p$ are corresponding quasimomenta and spin
projections are defined with respect to the magnetization
direction of a given half-space with opposite magnetization
directions (spin is supposed to be conserving in the "global"
coordinate system). For the tunneling current $I$ in the second
order in $T_{kp}$ one has
\begin{eqnarray}
I &\propto& \sum_{kqp}|T_{kp}|^2 \left[ 1+N_q -f(t_{p-q})\right] \nonumber \\
  &\times & \left[ f(t_k)-f(t_k+eV)\right] \delta(eV+t_k-t_{p-q}+\omega_q)
\end{eqnarray}
where $e$ is the electron charge and $V$ is the bias
\cite{ourtransport}. The differential conductance of the junction
with antiparallel magnetization directions is just proportional
(for zero temperature) to the density of the non-quasiparticle
states,
$$dI/dV \propto N_{NQP}(eV)$$

Note that the value $N_{NQP}(eV)$ vanishes for $|eV|< \hbar
\omega_0$, where $\hbar \omega_0$ is an anisotropy gap in the
magnon spectrum \cite{ourtransport}, which is small, but could be
changed by suitable substitution \cite{excipienti}. It is
worthwhile to note also that the nonquasiparticle states play an
important role (together with the classical effect of the smearing
of the gap by spin-disorder scattering) in the depolarization of
the states near the Fermi energy at finite temperatures; again,
estimations of the tunneling magnetoresistance based on a simple
Stoner-like picture of the energy spectrum appear to be completely
unrealistic \cite{berlin}.

\section{Non-quasiparticle states and electronic structure}
To perform the calculations we used the newly developed LDA+DMFT
scheme \cite{EMTODMFT}. This method is based on the so-called
Exact Muffin-tin Orbitals (EMTO) scheme \cite{andersen94,vitos00}
within a screened KKR \cite{weinberger90} approach, frozen core
together with the local spin density approximation (LDA). The
correlation effects are treated in the framework of dynamical mean
field theory (DMFT) \cite{Georges96}, with a spin-polarized
T-matrix Fluctuation Exchange (SPTF) type of DMFT solver
\cite{Katsnelson01}. The SPTF approximation is a multiband
spin-polarized generalization of the fluctuation exchange
approximation (FLEX) \cite{bicker89,Katsnelson99}, but with a
different treatment of particle-hole (PH) and particle-particle
(PP) channels. The particle-particle (PP) channel is described by
a $T$-matrix approach \cite{galitski63} giving a renormalization
of the effective interaction. This effective interaction is used
explicitly in the particle-hole channel. Justifications, further
developments and details of this scheme can be found in Ref.
\onlinecite{Katsnelson01}. The spin-polaron process is described
by the fluctuation potential matrix $W^{\sigma \sigma ^{\prime
}}(i\omega )$ with $\sigma=\pm$, defined in a similar way as in
the spin-polarized FLEX approximation \cite{Katsnelson99}:

\begin{equation}\label{W}
{\hat W}(i\omega )=\left(
\begin{array}{cc}
{W}^{++}(i\omega ) & {W}^{+-}(i\omega ) \\
{W}^{-+}(i\omega ) & {W}^{--}(i\omega )
\end{array}
\right).
\end{equation}

The essential feature here is that the potential (\ref{W}) is a
complex energy dependent matrix in spin pace with {\it
off-diagonal} elements:
\begin{equation}
W^{\sigma, -\sigma}(i\omega)=U^m(\chi^{\sigma, -\sigma}(i\omega )-
\chi_0^{\sigma, -\sigma}(i\omega ))U^m
\end{equation}
where $U^m$ represents the bare vertex matrix corresponding to the
transverse magnetic channel, $\chi^{\sigma, -\sigma}(i\omega )$ is
an effective transverse susceptibility matrix and $\chi^{\sigma,
-\sigma}_0(i\omega )$ is the bare transverse susceptibility
\cite{Katsnelson99}. $i\omega$ are fermionic Matsubara frequencies
and $(m)$ corresponds to the magnetic interaction channel
\cite{bicker89,Katsnelson99}. The local Green functions as well as
the electronic self-energies are spin diagonal for collinear
magnetic configurations. In this approximation the electronic
self-energy is calculated in terms of the effective interactions
in various channels. The particle-particle contribution to the
self-energy  was combined with the Hartree-Fock and the second
order contributions \cite{Katsnelson99}. To ensure a  physical
transparent description in the current implementation the combined
particle-particle selfenergy is presented by its a Hartree
$\Sigma^{(TH)}$ and Fock $\Sigma^{(TF)}$ types of contributions:
\begin{equation}
\Sigma = \Sigma^{(TH)}+ \Sigma^{(TF)} + \Sigma^{(ph)}
\end{equation}
where the particle-hole contribution $\Sigma^{(ph)}$ reads:
\begin{equation}\label{selfph}
\Sigma_{12 \sigma} ^{(ph)} = \sum_{34 \sigma^{\prime}} W_{1342}^{\sigma
\sigma^{\prime}} G_{34}^{\sigma^{\prime}}
\end{equation}

Since some part of the correlation effects are included already in
the local spin-density approximation (LSDA) ``double counted''
terms should be taken into account. To this aim, we start with the
LSDA electronic structure and replace $\Sigma_{\sigma}(E)$ by
$\Sigma_{\sigma}(E)-\Sigma_{\sigma}(0)$ in all equations of the
LDA+DMFT method \cite{FeNi}, the energy $E$ being relative to the
Fermi energy and $E_F=0$. It means that we only add {\it
dynamical} correlation effects to the LSDA method.

In our calculations we considered the standard representation of
the zinc-blende structure with an fcc unit cell containing four
atoms: Cr$(0,0,0)$, As$(1/4,1/4,1/4)$ and two vacant sites
E$(1/2,1/2,1/2)$ and E1$(3/4,3/4,3/4)$. We used the experimental
lattice constant of GaAs($5.65\AA$) and InAs($6.06\AA$) for all
the calculations, and the equilibrium value for the bulk CrAs
(a$_{eq}=5.8\AA$), obtained by FLAPW \cite{Shirai01}. To calculate
the charge density we integrate along a contour on the complex
energy plane which extends from the bottom of the band up to the
Fermi level \cite{vitos00}, using 30 energy points. For the
Brillouin zone integration we sum up a k-space grid of 89 points
in the irreducible part of the Brillouin zone. A cutoff of
$l_{max}=8$ for the multi-pole expansion of the charge density and
a cutoff of $l_{max}=3$ for the wave functions was used. The
Perdew-Wang \cite{PW} parameterization of the LDA to the exchange
correlation potential was used.

\section{Results and discussions}
The LDA+DMFT calculations were carried out for three lattice
constants: the GaAs, InAs and the "equilibrium" one. The
corresponding LDA computational results agree with previous ones
\cite{Akinaga00,Shirai01,Galanakis03}. In order to evaluate the
average Coulomb interaction on the Cr atoms and the corresponding
exchange interactions we start with the constrained LDA method
\cite{AnisimovU}. In our case the constrained LDA calculation
indicates that the average Coulomb interaction between the Cr $3d$
electrons is about $U=6.5$ eV with an exchange interaction energy
about $J=0.9$ eV. It is important to note that the values of the
average Coulomb interaction parameter slightly decrease going from
the GaAs ($U=6.59$ eV) to InAs ($U=6.25$ eV) lattice constants, 
see table Tab. \ref{tab1}.
This is in agreement with a naive picture of a less effective
screening due to increasing of the distances between the atoms.

The typical insulating screening used in the constraint
calculation \cite{AnisimovU} should be replaced by a metallic kind
of screening. The metallic screening will lead to a smaller value
of $U$. Unfortunately, there are no reliable schemes to calculate
$U$ in metals, therefore we choose some intermediate value of
$U=2$ eV and $J=0.9$ eV. It is important to realize that there are
no significant changes in the values of the average Coulomb
interaction for the studied lattice structures, and that the
exchange interaction is practically constant. Note that physical
results are not very sensitive to the value of $U$, as it was
demonstrated by us for NiMnSb \cite{NQP}.

\begin{figure}[h]
\centerline{\psfig{figure=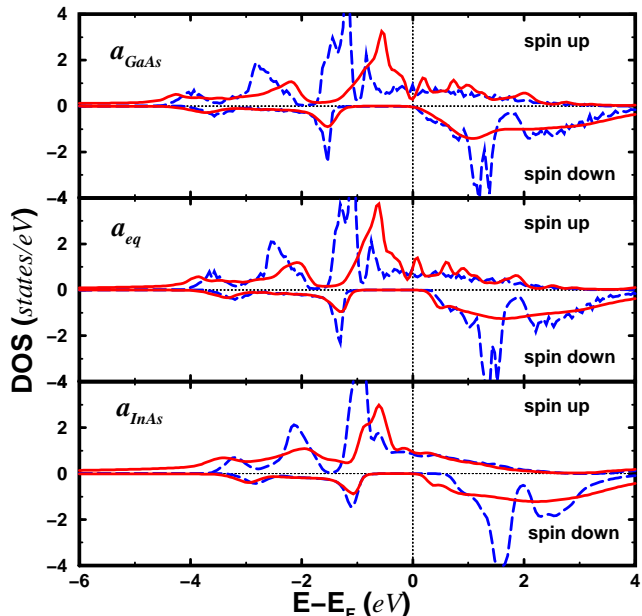,height=3.0in}}
\caption{Cr density of states calculated in LSDA (dashed line) and LSDA+DMFT (solid line)
methods corresponding to a temperature of $T=200K$, average Coulomb interaction
parameter $U=2$ eV and exchange $J=0.9$ eV. The non-quasiparticle states are clearly
visible for lattice parameters larger than $a_{eq}=5.8\AA$, in the unoccupied part
for minority spin channel just above the Fermi level, around $0.5eV$.}
\label{200_cras_DOS}
\end{figure}

The LSDA and LSDA+DMFT calculation for the density of states, DOS,
is presented in Fig. \ref{200_cras_DOS}. Depending on the lattice
constant the Cr and As atoms loose electrons and this charge is
gained by the vacant sites. As a result the Fermi level is moving
from the right edge of the gap as for the case of GaAs substrate,
towards the middle of the gap for a InAs substrate. The Cr moments
are well localized due to a mechanism being similar to that of
localization of the magnetic moment on the Mn atom in the Heusler
NiMnSb \cite{deGroot83}. Note that the local Cr spin moment is
more than $3 \mu_B$, the As atom possesses a small induced
magnetic moment (of order of $-0.3 \mu_B$) coupled antiparallel to
the Cr one. The results are presented in table \ref{tab2}.
Calculated, DMFT, Cr magnetic moments increased in comparison with
the LDA results, due to the localization tendency of the Cr$-3d$
states as a consequence of correlation effects \cite{EMTODMFT}.

According to our calculations, the system remains half-metallic
with a pretty large band gap (about $1.8$ eV) for all the lattice
constants compared with the band gap of the NiMnSb  which is only
$0.75$ eV \cite{deGroot83}.

In Fig. \ref{200_cras_DOS} the non-quasiparticle states are
visible for lattice parameter higher than the equilibrium one,
with a considerable spectral weight in the case of the InAs
substrate. InAs is widely used as substrate material for the
growth of different compounds. According  to our calculations this
situation is very favorable for the experimental investigation of
the formation of non-quasiparticle states as well as the
investigation of their nature.

Comparing with the case of the GaAs lattice parameter, for the
InAs substrate, the most significant change in the electronic
structure is suffered by the  As-$p$ states. Having a larger
lattice constant the Cr atom acquires a slightly larger magnetic
moment. In the LDA calculations, nevertheless, the magnetic moment
per unit cell is integer $3 \mu_B$. Expanding the lattice constant
from GaAs to InAs lattice  the Cr states become more atomic like
and therefore the spin magnetic moment increases. This is
reflected equally in the charge transfer which is smaller for the
InAs lattice parameters. A larger Cr moment induces a large spin
polarization of the As-$p$ states, compensating the smaller $p-d$
hybridization, the total moment remaining at its integer value of
$3 \mu_B$ \cite{Galanakis03}.

\begin{figure}[h]
\centerline{\psfig{figure=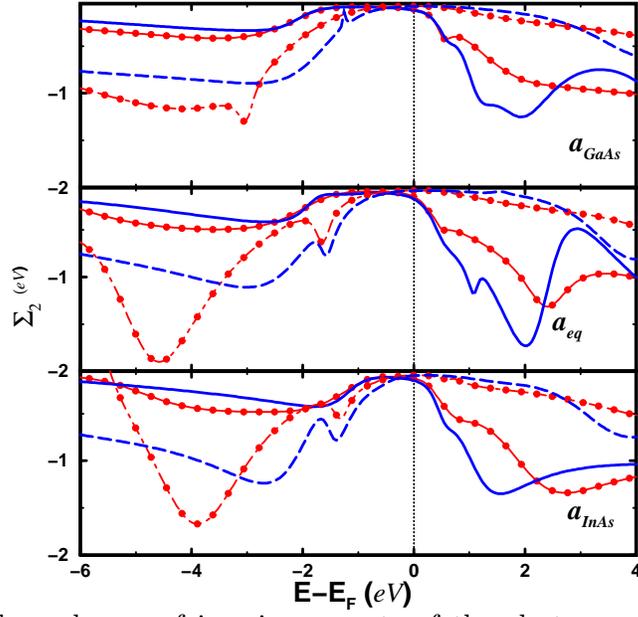,height=3.0in}}
\caption{Energy dependences of imaginary parts of the electron self-energy $\Sigma_2(E)$, for
lattice constants of GaAs (a), equilibrium one (b), and of InAs (c): $e_g$ down solid line,
$t_{2g}$ down decorated solid line,  $e_g$ up dashed line, $t_{2g}$ up decorated dashed line.}
\label{sigma}
\end{figure}

Note that the essential difference of the many-body electronic
structure for the lattice constants of GaAs and InAs is completely
due to the difference in the position of the Fermi energy with
respect to the minority-spin band gap whereas the self-energy
characterizing the correlation effects is not changed too much
(Fig. \ref{sigma}). The total density of states $N\left( E\right)
$ is rather sensitive to the difference between the band edge
$E_{c}$ and the Fermi energy $E_{F}$. If this difference is very
small (i.e., the system is close to the electronic topological
transition $E_{c}\rightarrow E_{F}$) one can use a simple
expression for the singular contribution to the bare density of
states, $\delta N_{0}\left( E\right) \propto \sqrt{E-E_{c}}$
$\left( E>E_{c}\right)$. The appearance of the complex self-energy
$\Sigma \left( E\right) =\Sigma _{1}\left( E\right) -i\Sigma
_{2}\left( E\right) $ \ changes the singular contribution as
\begin{equation}
\delta N\left( E\right) \propto [ \sqrt{ Z_{1}^{2}\left(E\right)+
\Sigma_{2}^{2}\left(E\right) }  + Z_{1}\left(E\right) ]^{1/2}
\end{equation}
where $Z_1\left(E\right)=E-E_c-\Sigma_1 \left(E\right)$ (cf. Ref.
\onlinecite{ETT}). Assuming that the self-energy is small in
comparison with $E-E_{c}$ one can find for the states in the gap
$\delta N\left( E\right) \propto \Sigma _{2}\left( E\right)
/\sqrt{E_{c}-E}$ $\left( E<E_{c}\right) .$ One can see that the
shift of the gap edge changes drastically the density of states
for the same $\Sigma _{2}\left( E\right).$

In the remaining part of the paper we elaborate more on the
possibility of the practical use of tunable properties of
non-quasiparticle states in CrAs grown on different substrates.
For most of the applications room-temperature and the stability of
the ferromagnetic state are the most important prerequisites. The
ferromagnetic CrAs material might be grown on III-V semiconductors
similarly to the zinc-blende CrSb \cite{Zhao01}. We observed the
presence of the NQP states for CrAs lattice parameters larger than
$5.8 \AA$. It was found experimentally that at 300$K$, around this
value of the lattice parameter a stable solid solution of
Ga$_{0.65}$In$_{0.35}$As is formed \cite{Harland66}. So, from the
practical point of view a $65\%$ of gallium in a
Ga$_{x}$In$_{1-x}$As compound would constitute the ideal substrate
on which the CrAs half-metal would present tunable properties of
its NQPs. Having the Ga$_{0.65}$In$_{0.35}$As substrate CrAs could
be a part of the epitaxial III-V structure providing an easy way
to integrate with the existing semiconductor technology.

In addition, to determine possible substrates for growth of
layered half-metallic materials, electronic structure calculations
were carried out for lattice parameters in the range $5.60\AA$ to
$6.03\AA$ \cite{Fong04}. According to these calculations \cite{Fong04}, growth
with minimal strain might be accomplished in a half-metallic
multilayer system grown on InAs substrate which would be the
best choice to evidence the NQP states, since the Fermi energy 
is situated in this case far
enough from the bottom of the conduction band.

The existence of NQP states is a manifestation of the many-body
interaction at finite temperatures, magnetoelectronic applications
being able to measure their presence.

\begin{figure}[h]
\centerline{\psfig{figure=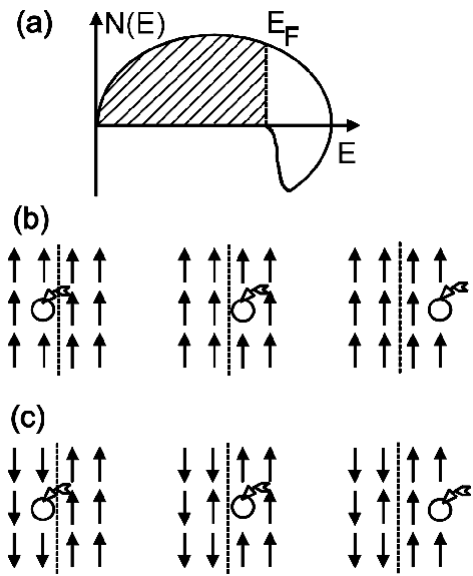,height=3.15in}}
\caption{The tunneling transport between strongly correlated ferromagnets. The density
of states in the lower Hubbard band (a) is provided by standard current states
for majority-spin electrons (above) and by non-quasiparticle states for minority-spin
electrons (below), the latter contribution being non-zero only above the Fermi energy
(occupied states are shadowed). However, the tunneling is possible both for parallel (b)
and antiparallel (c) magnetization directions.}
\label{model}
\end{figure}

A very high sensitivity of the minority-electron density of states
near the Fermi energy to the lattice constant opens a new
interesting opportunity to run the current in CrAs-based tunnel
junction. Suppose we have an antiparallel orientation of the
magnetizations in the tunnel junction (such as shown in Fig.
\ref{model}c), then the $I-V$ characteristic is determined by the
density of the non-quasiparticle states. Thus if we will influence
the lattice constant (e.g., using a piezoelectric material) one
can modify the differential conductivity. This makes CrAs a very
promising material with tunable characteristics which opens new
ways for applications in spintronics.

This work is part of the research programme of the
Stichting voor Fundamenteel Onderzoek der Materie (FOM),
which is financially supported by the
Nederlandse Organisatie voor Wetenschappelijk Onderzoek (NWO).

\begin{table}[h]
\begin{tabular}{c|cccc|cc}
& Cr & As & E & E1 & $U$ & $J$ \\
& $(a.u)$ & $(a.u)$ & $(a.u)$ & $(a.u)$ &
$(eV)$ & $(eV)$ \\ \hline\hline
$R_{MT}^{GaAs} $ & 2.565 & 2.688  & 2.565  & 2.688 & 6.59 & 0.93  \\
$R_{MT}^{eq.} $ & 2.633 & 2.759  & 2.633  & 2.759 & 6.46 & 0.93  \\
$R_{MT}^{InAs} $ & 2.751 & 2.883  & 2.751  & 2.883 & 6.25 & 0.93
\end{tabular}
\vspace{0.25cm}
\caption{The muffin-tin radii of the Cr, As and the two types of empty spheres
placed as vacancies of the ZB structure. The constreined LDA
values for the Coulomb and exchange interactions are presented.}
\label{tab1}
\end{table}

\begin{table}[tbp]
\begin{tabular}{c|ccccc|ccc}
& Cr & As & E & E1 &  Total & $T$ & $U$ & $J$ \\
& $(\mu_B)$ & $(\mu_B)$ & $(\mu_B)$ & $(\mu_B)$ & $(\mu_B)$
& $(K)$ & $(eV)$ & $(eV)$ \\ \hline\hline
$\mu_{LDA}^{GaAs} $ & 3.191 & -0.270 & -0.009 & 0.089 & 3.00 & -  & - & -  \\
$\mu_{DMFT}^{GaAs}$ & 3.224 & -0.267 & -0.023 & 0.067 & 3.00 & 200 & 2 & 0.9 \\  \hline
$\mu_{LDA}^{eq.} $ & 3.284 & -0.341 & -0.018 & 0.076 & 3.00 & -  & - & -  \\
$\mu_{DMFT}^{eq.}$ & 3.290 & -0.327 & -0.024 & 0.068 & 3.00 & 200 & 2 & 0.9 \\ \hline
$\mu_{LDA}^{InAs} $ & 3.376 & -0.416 & -0.025 & 0.066 & 3.00 & -  & - & -  \\
$\mu_{DMFT}^{InAs}$ & 3.430 & -0.433 & -0.033 & 0.043 & 3.00 & 200 & 2 & 0.9
\end{tabular}
\vspace{0.25cm}
\caption{Summarizing table containing the results of our calculation. CrAs magnetic moments
corresponding to the GaAs, InAs and the equlibrium lattice
constant $a_{eq}$. For the later one the value $a_{eq}=5.8\AA$ was used, Ref. [4].
Parameters of the DMFT calculations are presented in the last three columns
of the table.}
\label{tab2}
\end{table}

\end{document}